# Generalized coherent wave control at dynamic interfaces

Youxiu Yu, Dongliang Gao[*], Yukun Yang, Liangliang Liu, Zhuo Li, Qianru Yang, Haotian Wu, Linyang Zou, Xiao Lin, Jiang Xiong, Songyan Hou, Lei Gao[*], and Hao Hu[*]

Y. Yu, D. Gao, L. Gao
School of Physical Science and Technology & Jiangsu Key Laboratory of Thin Films
Soochow University, Suzhou 215006, China
E-mail: dlgao@suda.edu.cn

Y. Yang, L. Liu, Z. Li, H. Hu
National Key Laboratory of Microwave Photonics & College of Electronic and Information Engineering,
Nanjing University of Aeronautics and Astronautics, Nanjing 211106, China.
E-mail: hao.hu@nuaa.edu.cn

Q. Yang, H. Wu, L. Zou
School of Electrical and Electronic Engineering
Nanyang Technological University, Singapore 639798, Singapore.

X. Lin
Interdisciplinary Center for Quantum Information, State Key Laboratory of Modern Optical Instrumentation
Zhejiang University, Hangzhou 310027, China

J. Xiong
Computational Electromagnetics Laboratory, Institute of Applied Physics
University of Electronic Science and Technology of China, Chengdu 610054, China.

S. Hou
Guangzhou Institute of Technology
Xidian University, Guangzhou 510555, China.

L. Gao

School of Optical and Electronic Information

Suzhou City University, Suzhou 215104, China.

E-mail: leigao@suda.edu.cn

L. Gao

School of Optical and Electronic Information

Suzhou City University, Suzhou 215104, China.

E-mail: leigao@suda.edu.cn

Coherent wave control is of key importance across a broad range of fields such as electromagnetics, photonics, and acoustics. It enables us to amplify or suppress the outgoing waves via engineering amplitudes and phases of multiple incidences. However, within a purely spatially (temporally) engineered medium, coherent wave control requires the frequency of the associated incidences to be identical (opposite). In this work, we break this conventional constraint by generalizing coherent wave control into a spatiotemporally engineered medium, i.e., the system featuring a dynamic interface. Owing to the broken translational symmetry in space and time, both the subluminal and superluminal interfaces allow interference between scattered waves regardless of their different frequencies and wavevectors. Hence, one can flexibly eliminate the backward- or forward-propagating waves scattered from the dynamic interfaces by controlling the incident amplitudes and phases. Our work not only presents a generalized way for reshaping arbitrary waveforms but also provides a promising paradigm to generate ultrafast pulses using low-frequency signals. We have also implemented suppression of forward-propagating waves in microstrip transmission lines with fast photodiode switches.

## 1. Introduction

Interaction between waves occurs when the orthogonality between two or more incidences is broken by the spatial structures.[1] The energies of the outgoing waves scattered from the structures stringently depend on the relative amplitudes and phases of these incidences. By precisely controlling the relative amplitudes and phases of multiple incidences, the behaviors of outgoing waves are artificially engineered. This phenomenon is referred as the coherent wave control.[2, 3] Coherent wave control is currently a very active research field as it leads to a wealth of extraordinary effects,[4] among which coherent perfect absorption is perhaps the most striking phenomenon.[5, 6] Under the conditions of coherent perfect absorption, e.g., introducing a precise amount of dissipation into an object, the electromagnetic fields that are initially scattered off will be completely absorbed by the object at a particular frequency.[7, 8] Moreover, in the nanostructured waveguides, coherent wave control is useful for the selective excitation of bulk or edge states.[9] These remarkable effects facilitate many practical applications in optical absorbers, interferometers, ultrafast optical devices, laser cooling, and atomic clocks.[2, 10-14]

In the recent past, time-varying materials, whose material properties vary rapidly in time instead of space,[15, 16] have received close attention, owing to their rich physics and many counterintuitive phenomena. Specifically, a wealth of papers have already demonstrated that temporally engineered medium is a promising platform to achieve the advanced electromagnetic wave manipulation, such as magnetic-free nonreciprocity,[17] double-slit time diffraction,[18] and stationary charge radiation.[19] These novel effects render emerging applications, such as antireflection temporal coatings,[20] non-resonant lasers,[21] and analog computing.[22]

Inspired by the applications in spatially engineered media, the concept of coherent wave control has been utilized in its temporally engineered counterparts. Very recently, coherent wave control has been experimentally demonstrated in a system featuring a temporal interface.[23] In this experiment, the strong wave interaction occurs only in the time scale of several nanoseconds, allowing sculpting light with light in an instantaneous and ultra-broadband manner.

However, identical (opposite) frequency of the incidences is required to ensure that co-scattering waves have the same frequencies and momenta for realizing coherent wave control in purely spatially (temporally) engineered media.[24] To be specific, due to the time translational symmetry, the interference in purely spatially engineered media necessitates all incidences to be of the same frequencies. On the other hand, as the temporally engineered media follow the spatial translation symmetry, the interference occurs only when all incidences possess identical momenta but opposite frequencies. Hence, the same magnitude of incidence frequencies is always a necessary requirement to realize coherent wave control at the purely spatial or temporal interface. This constraint fundamentally prevents potential applications of coherent wave control in tailoring high-frequency waves with low-frequency ones. To date, how to achieve coherent wave control via multiple incidences with arbitrary frequencies remains unknown.

In this work, we propose a generalized coherent wave control within the systems with dynamic interfaces, i.e., spatiotemporally engineered media whose refractive indices undergo instantaneous changes in both the spatial and temporal domains.[16, 25] With the growing attention to spatiotemporally engineered media,[26-28] plenty of physical concepts in conventional materials have been extended to this system, e.g., quarter-wave impedance transformers,[29] Cherenkov radiation in the vacuum,[30] and Talbot effect.[31] Enlightened by recent advances, we have shown that spatiotemporally engineered media provide an indispensable way to realize coherent wave control with different magnitudes of incident frequencies. The underlying mechanism is that the wave interaction process is no longer constrained by momentum and energy conservation rules. Because both the time and space translational symmetries are simultaneously broken due to the space-time interface. Moreover, the moving interface introduces an extra degree of freedom, enabling flexible frequency and wavevector transitions for incident waves. Even when two counter-propagating incidences have entirely different frequency amplitudes, outgoing waves could still have the same frequencies and wavevectors simply by controlling the speed of the moving interface. Based on the generalized coherent wave control, we demonstrate the powerful capability of generalized coherent control in eliminating space-time reflection/transmission, reshaping arbitrary waveforms and producing new signals. In addition, we demonstrate generalized coherent control in microstrip transmission lines with fast photodiode switches.

## 2. Principle

We consider the coherent modulation between two counter-propagating waves with different frequencies at a dynamic interface. The dynamic interface system consists of two media with refractive indices of $n_1$ and $n_2$, respectively. Depending on the interface velocity $v$, the dynamic interfaces can be classified into subluminal-interface ($|\beta| < \min\{1/n_1, 1/n_2\}$) and superluminal-interface ($|\beta| > \max\{1/n_1, 1/n_2\}$), where $\beta$ is the interface velocity normalized to the speed of light in the vacuum (i.e., $\beta = v/c$).[16] The frequencies and wavevectors are denoted as $\omega_{lm}^{i,\pm}$ and $k_{lm}^{i,\pm}$. Here, $l = f, b$ refer to forward and backward incidences, respectively; $m = 1, 2$ refer to media with refractive indices of $n_1$ and $n_2$, respectively; $i$, $+$, and $-$ correspond to the incident, transmitted, and reflected waves, respectively. The phase difference between the two counter-propagating waves is denoted as $\varphi$, and the contrast between the amplitudes of backward and forward incidences is denoted as $A$. Both the interface and incidences propagate only along the $x$-dimension.

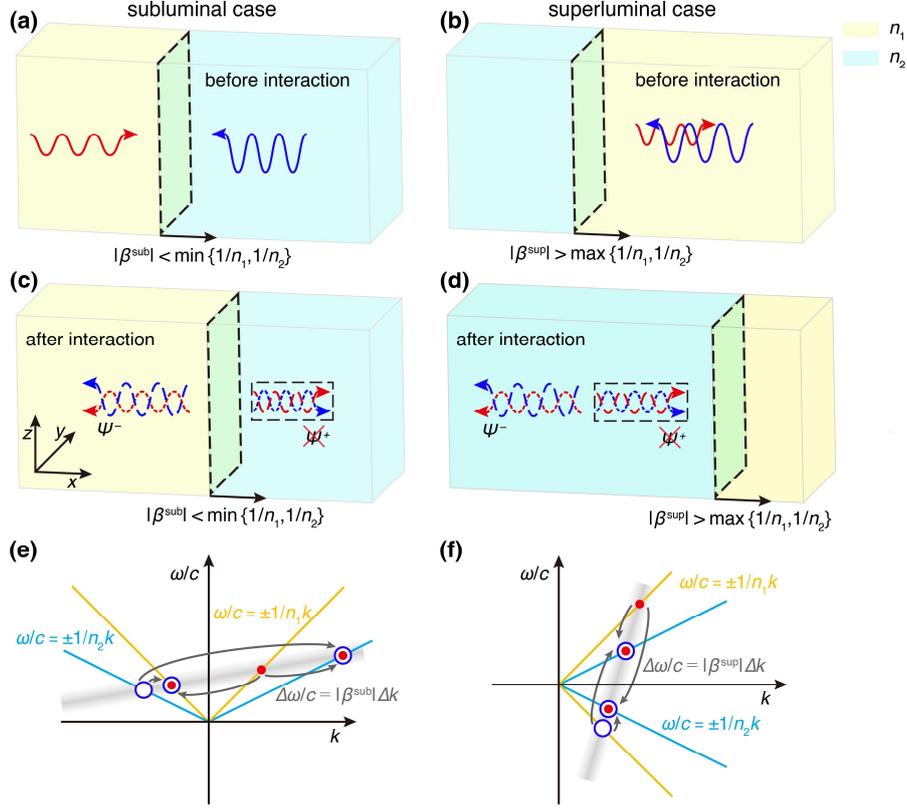

**Figure 1.** Conceptual demonstration of generalized coherent wave control at the dynamic interface. (a, b) Distribution of two counter-propagating waves in media before the interaction with subluminal and superluminal interfaces, respectively. (c, d) Distribution of two counter-propagating waves in media after the interaction with subluminal and superluminal interfaces, respectively. (e, f) Schematic of mode transitions in the momentum-energy space for subluminal and superluminal regimes, respectively. In (a-d), the incident, reflected, and transmitted waves are marked in solid, short-dashed, and long-dashed lines, respectively. In all the panels, forward and backward waves are marked in red and blue, respectively.

We first establish the condition for destructive interference of electromagnetic waves at the subluminal interface. In the subluminal regime, the scattering behaviors of two counter-propagating waves are illustrated in **Figure 1**a, c, where the reflected and transmitted waves propagate at different sides of the dynamic interface. The red (blue) solid, short-dashed, and long-dashed lines represent the forward (backward) incident wave and its reflected and transmitted waves, respectively. Wave interaction occurs only when the outgoing waves are propagating in the same direction share equivalent frequencies, i.e., the transmitted wave of the forward incident wave and the reflected wave of the backward incident wave have identical frequencies, and vice versa. Considering the extra energy and momentum provided by the dynamic interface for incidences, the frequencies of the two counter-propagating incident waves and the interface velocity must satisfy:

$$\omega_{b2}^i = \frac{1-n_1\beta}{1+n_2\beta}\omega_{f1}^i. \qquad (1)$$

The corresponding mode transitions in the subluminal regime are shown in Figure 1e. In the system of uniformly moving interface, the mode transition must satisfy $\Delta\omega/c = |\beta^{\text{sub}}|\Delta k$, where $\Delta\omega$ and $\Delta k$ refer to the variations of frequency and wavevector, respectively. After the two incident waves with different frequencies interact with the space-time interface with suitable velocity $\beta^{\text{sub}}$, the frequencies and wavevectors of the forward- (backward-) propagating wave from forward incident wave (as remarked by red circles) become identical to those of backward- (forward-) propagating wave from backward incident wave (as remarked by blue circles). The identical frequencies and wavevectors enable interference between scattering waves in the subluminal regime. The diagram of mode transition in the superluminal regime is displayed in Figure 1f, similar to that of subluminal regime. Due to the broken translational symmetry in the temporal and spatial domain [25, 29], the generalized coherent wave control no longer requires the incident frequencies to be identical. What's more, the conditions for coherent wave control, such as the relationship between incident frequencies, can be arbitrarily engineered by changing the interface velocity. Since the continuous field quantity at the uniformly-moving interface is $H$-$vD$, we set this parameter to be the wavefunction $\psi$. To further eliminate the outgoing waves propagating in a particular direction, e.g., the forward direction, the wavefunction of the following forward-outgoing wave should be fulfilled: $\psi^{\text{sub},+} = H^{\text{sub},+} - vD^{\text{sub},+} = T_f^{\text{sub}}(H_{zf}^+ - v_f D_{yf}^+) + R_b^{\text{sub}}(H_{zb}^- - v_b D_{yb}^-) = 0$, where the magnetic field and electric displacement of the scattering wave are $H_{zf}^+ = e^{ik_{f2}^+ x - i\omega_{f2}^+ t}$, $D_{yf}^+ = n_2 e^{ik_{f2}^+ x - i\omega_{f2}^+ t}/c$, $H_{zb}^- = A e^{i k_{b2}^- x - i \omega_{b2}^- t + i\varphi}$, $D_{yb}^- = A n_2 e^{i k_{b2}^- x - i \omega_{b2}^- t + i\varphi} / c$; $k_{f2}^+$ ($k_{b2}^-$) and $\omega_{f2}^+$ ($\omega_{b2}^-$) respectively refer to the wavevector and frequency of the forward- (backward-) propagating wave for forward (backward) incident wave; $A$ ($\varphi$) refers to the relative amplitude (phase) of the backward incident wave; $T_f^{\text{sub}}$ ($R_b^{\text{sub}}$) refers to the transmission (reflection) coefficient of the forward- (backward-) outgoing wave, respectively. Enforcing the boundary condition at the interface $x = \beta c t$, one can solve for the relative amplitude $A$ and phase $\varphi$ of the incidences. More details are provided in Section S1 of the Supporting Information. To realize the destructive interference of forward waves at the subluminal interface, the required amplitude contrast and phase difference of incidences are derived as

$$A = -\frac{2 n_2 (1 - n_1 \beta)}{(n_1 - n_2)(1 + n_2 \beta)} \tag{2}$$

$$\varphi = 0. \tag{3}$$

For backward waves, the derived condition of destructive interference in the subluminal regime is given in Section S1 of the Supporting Information. Note that we select $H - vD$ as the wavefunction $\psi$ in this work because of its continuous nature at the uniformly-moving interface.

Next, we investigate the condition for achieving destructive interference of electromagnetic waves at the superluminal interface. As shown in Figure 1b, d, the reflected and transmitted waves propagate at the same sides of the dynamic interface in the superluminal regime. The reflection and transmission coefficients of outgoing waves are now determined as shown in Table S2 in Section S1 of the Supporting Information. To achieve the wave interference, the incident frequencies in the superluminal regime are governed by

$$\omega_{b1}^i = \frac{1 - n_1 \beta}{1 + n_1 \beta} \omega_{f1}^i \tag{4}$$

and this relation can also be tuned by the interface velocity. The slight difference between Equation 1 and 4 results from the fact that two incidences are counter-propagating in the distinct media in the subluminal regime while those waves are propagating in the same media in the superluminal regime (see mode transition diagram in Figure 1e, f). Owing to this distinction, the condition for achieving destructive interference is modified as

$$A = -\frac{(n_1 + n_2)(1 - n_1\beta)}{(n_1 - n_2)(1 + n_1\beta)} \tag{5}$$

$$\varphi = 0. \tag{6}$$

## 3. Result and Discussion

To visualize the above-derived conditions for destructive interference, we demonstrate the spectra for forward- and backward-outgoing waves. Without loss of generality, the refractive indices of the media on both sides of the dynamic interface are set to be $n_1 = 1$, and $n_2 = 2$. The field intensity as a function of the interface velocity $\beta$ and relative amplitude $A$ is illustrated in **Figure 2**, where the relative phase difference is fixed at zero. The destructive interference (red dashed lines) is always observed at the dynamic interface regardless of its subluminal or superluminal nature. Equation 2, 3, 5, 6 and Figure 2 consistently show that different velocity of dynamic interface requires different relative amplitudes and phases of incidences to realize the elimination of outgoing waves. For example, we set $A = 2.29$ ($A = -0.29$) and $\varphi = 0$ when $\beta = 0.2$ to destructively interfere with the forward- (backward-) outgoing waves in the subluminal regime; and $A = -0.27$ ($A = -0.03$) and $\varphi = 0$ when $\beta = 1.2$ to destructively interfere with the forward- (backward-) outgoing waves in the superluminal regime. Notably, the destructive interference of backward-outgoing waves could be switched to that of forward-outgoing waves with a simple adjustment of the relative amplitude $A$. This property makes the generalized coherent wave control more flexible to manipulate outgoing waves, as compared to previously established techniques such as a temporal version of the Brewster effect[32] and a spatiotemporal coating[29], by which elimination of forward-outgoing waves is rather challenging. Significantly, in contrast to conventional coherent wave control, this finding provides a unique platform for coherent wave control without resorting to the identical magnitude of incident frequencies and wavevectors.

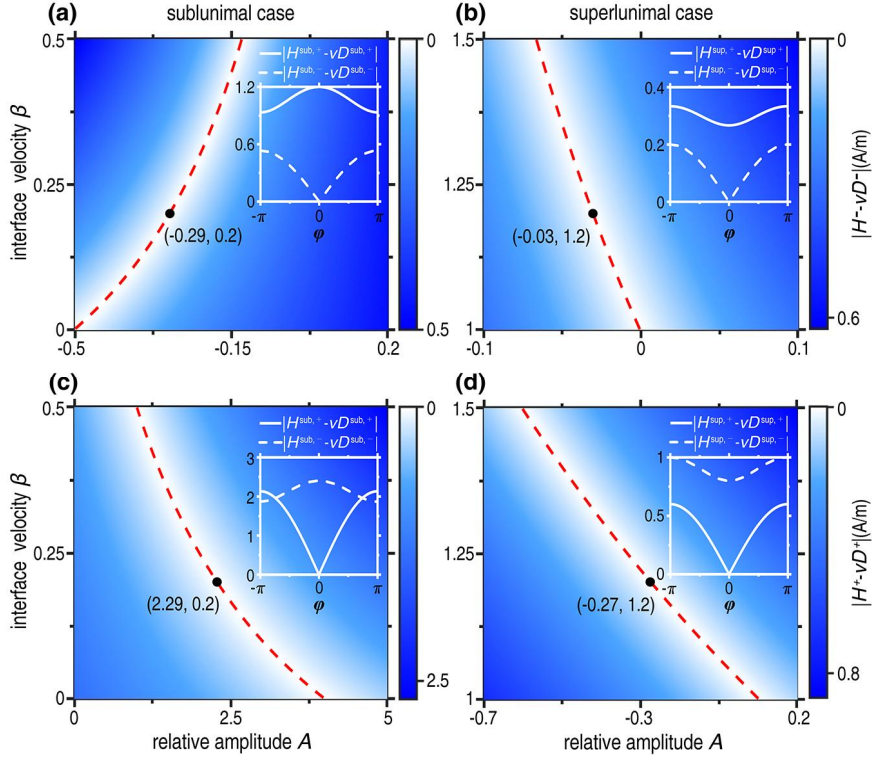

**Figure 2.** Influence of interface velocity, incident relative amplitude, and incident phase on the intensity of outgoing waves. (a, b) The intensity of backward-outgoing wave as a function of interface velocity and incident relative amplitude in the subluminal and superluminal regimes, respectively. (c, d) The intensity of forward-outgoing wave as a function of interface velocity and incident relative amplitude in the subluminal and superluminal regimes. In all the panels, we fix the phase difference of two incidences at $\varphi = 0$. The insets plot the intensity of outgoing waves as a function of phase difference when the $A$ is $-0.29$ (a), $-0.03$ (b), $2.29$ (c), and $-0.27$ (d), respectively. For the insets in (a & c), $\beta = 0.2$, and for the insets in (b & d), $\beta = 1.2$. The forward- and backward-outgoing waves are highlighted in white-solid and dashed lines, respectively. In (a-d), the red-dashed lines are obtained from Equation 2, 5, S1.2 and S1.3 at $\varphi = 0$. In all the panels here and below, we fix the refractive indices $n_1 = 1$, and $n_2 = 2$, and the wavelength of the forward incident wave at 1 μm (i.e., a frequency of $1.88 \times 10^{15}$ Hz), while the frequencies of the backward incident wave change with the interface velocity as shown in Figure S2.

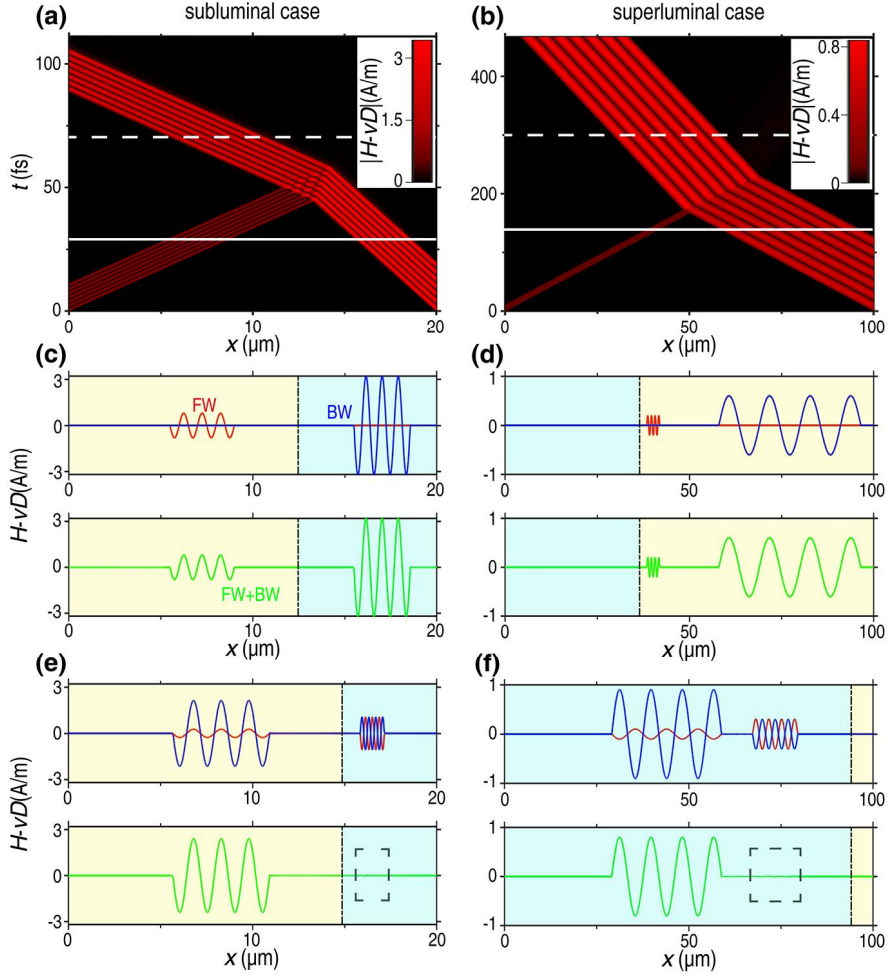

**Figure 3.** Eliminating forward-propagating waves with coherent wave control at dynamic interfaces. (a, b) Spatiotemporal field distributions in the subluminal and superluminal regimes, respectively. (c, e) Spatial field distributions in the subluminal regime at $t=30\,fs/70\,fs$ (as indicated by the white lines in (a)). (d, f) Spatial field distributions in the superluminal regime at $t=140\,fs/300\,fs$ (as indicated by the white lines in (b)). In (a, c, e), $\omega_{b2}^i = 4/7\omega_{f1}^i$, $A=2.29$, and $\beta=0.2$. In (b, d, f), $\omega_{b1}^i = -1/11\omega_{f1}^i$, $A=-0.27$, and $\beta=1.2$. In all the panels here and below, the dynamic interface is highlighted with black dashed lines.

To verify the performance of generalized coherent wave control, we implement the finite-difference time-domain (FDTD) numerical simulation. **Figure 3**a, b illustrates simulated field distributions in the spatiotemporal domain at subluminal and superluminal interface velocities, respectively. Note that generalized coherent wave control applies not only to monochromatic waves but also to pulse-like signals (only if spectra of incidences feature the similarity, seeing detailed explanation in Section S2 of the Supporting Information) In the subluminal regime with $\beta = 0.2$ (superluminal regime with $\beta = 1.2$), the central frequencies and relative amplitude of incident pulses are set as $\omega_{b2}^i = 4/7\omega_{f1}^i$ ($\omega_{b1}^i = -1/11\omega_{f1}^i$) and $A = 2.29$ ($A = -0.27$), respectively. Note that the negative frequency of the backward incident wave is a direct consequence when enforcing the frequency condition (i.e., Equation 4) in the superluminal regime. The above parameters fulfill the conditions for the destructive interference of forward-outgoing waves. As a consequence, the forward-outgoing wave is no longer observed after the wave interaction at dynamic interfaces. To further illustrate this effect, we show the field distributions at different time moments in Figure 3c-f. In the subluminal regime, the forward and backward incidences reside on different sides of the dynamic interface at $t = 30\,fs$ (i.e., before the wave interaction). At $t = 70\,fs$ (i.e., after the wave interaction), the field distribution shows that the forward-outgoing wave is negligible on the right side of the dynamic interface due to the destructive interference between the transmitted wave from forward incidence and the reflected wave from backward incidence. A similar effect is also observed in the superluminal regime, where the forward-outgoing wave disappears on the right side of the dynamic interface. Regarding the energy transfer, the moving interface initially providing energy toward the backward-outgoing wave is gradually switched to the one absorbing energy from the incident wave, as the interface velocity increases. More discussions on the destructive interference of backward-outgoing waves and energy are present in Section S2 of the Supporting Information. The generalized coherent wave control also exhibits attractive applications in reshaping waveforms (as see Figure S6). In contrast to the temporal counterpart,[23] the same frequency or wavevector of the incident wave is not required in this scenario. A more detailed discussion is available in Section S5 of the Supporting Information.

The generalized coherent wave control offers a novel approach for producing ultrafast pulses characterized by both short durations and extremely high frequencies, which are crucial for investigating ultrafast science and manipulating matter at the atomic and molecular levels.[33, 34] As is shown in **Figure 4**, the numerical simulations illustrate this concept: a forward-propagating continuous incidence is interacting with a train of backward-propagating idler pulses at a low frequency (see subluminal and superluminal regimes in Figure 4a, b, respectively). When the relative amplitude and phase of idler pulses satisfy the destructive interference condition, the transmitted signal is reshaped into a train of pulses with amplitudes comparable to that of incidences (see black dashed boxes in Figure 4c, d). Via proper control of the temporal separation between the neighboring idler pulses, the duration of the output signal could be flexibly engineered (which can be made down to 1.5 periods in Figure 4c, d). Due to the non-Hermitian nature of the system, the frequency of the produced pulse is 2.33 (1.57) times higher than that of idler pulses if the interface velocity is $\beta = 0.2$ ($\beta = 1.2$) in Figure 4c, d.

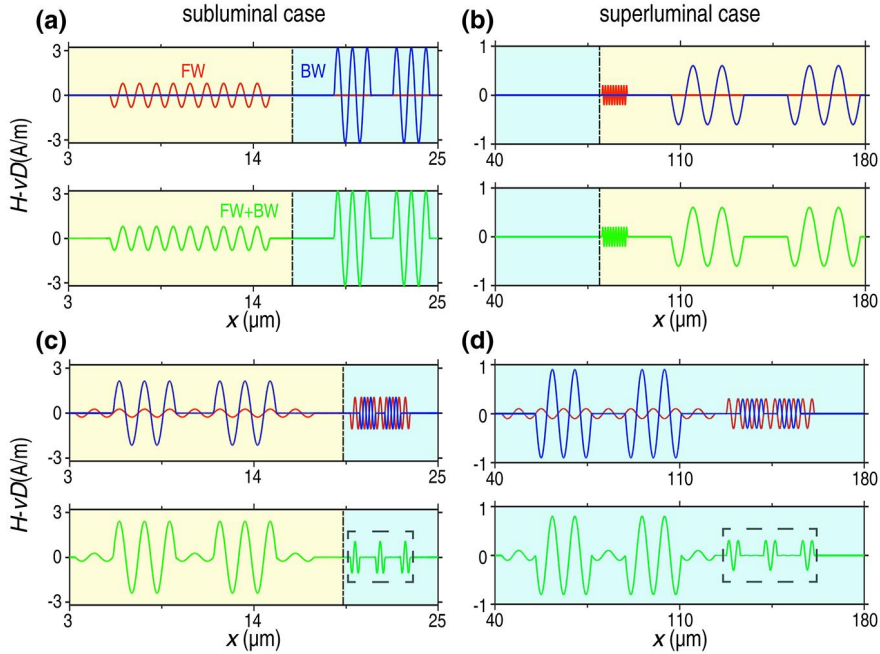

**Figure 4.** Generating ultrafast pulses with coherent wave control at the dynamic interface. (a, b) Spatial field distributions before the wave interaction with the subluminal and superluminal interfaces, respectively. (c, d) Spatial field distributions after the wave interaction with the subluminal and superluminal interfaces, respectively. In (a-d), the duration of the forward-propagating incident pulse is 9.5 periods, while that of the backward-propagating incident pulse is 2.5 periods. The temporal gap between two neighboring backward-propagating incident pulses is 1.5 periods.

## 4. Experiment Proposal

Inspired by recent experimental advances of coherent wave control at temporal interfaces,[23] we propose a platform of microstrip transmission lines (MTLs) to realize the generalized coherent wave control. **Figure 5**a illustrates an equivalent circuit model of a spatiotemporal MTL. The MTLs consist of 97 unit cells with the period of $\Delta x = 10$ mm. Each unit cell consists of a parallel capacitor ( $C_i = 2$pF, where $i = 1,2,3\cdots n$ ) in series with a fast photodiode. The on/off state of the capacitor is controlled by a fast photodiode acting as a switch. The details of the switch module and component parameters are provided in Section S6 of the Supporting Information. When the photodiode is not triggered (triggered) by the light, the switch is in the off (on) state and the equivalent refractive index of the MTLs is $n_1 = 2.60$ ( $n_2 = 3.80$ ). The triggering scheme of the photodiode is illustrated in Figure 5b: the photodiodes are triggered in sequence with the same time interval ( $\Delta t = 1$ ns ) starting from $t = 1$ ns. In this setup, the interface velocity is calculated as $\beta = -0.0334$, which is in the subluminal regime. When the frequency of the backward incidence is 500MHz, the frequency and amplitude of the forward incidence to eliminate forward-propagating waves are calculated as 401.7MHz and 7.9, respectively. Our simulation results from Advanced Design System (ADS) software are presented in Figure 5c, where the red and blue solid lines denote the signals probed at port 1 and port 2, respectively. The two counter-propagating signals interact with the dynamic interface at $t = 40$ ns. After interactions, signal at port 2 reflects that the forward-propagating wave generated at the dynamic interface (see the dashed rectangle in Figure 5c) is vanishing when the above condition to eliminate forward-propagating waves is completed. This phenomenon lays a solid foundation for the subsequent experimental realization of generalized coherent wave control at dynamic interfaces.

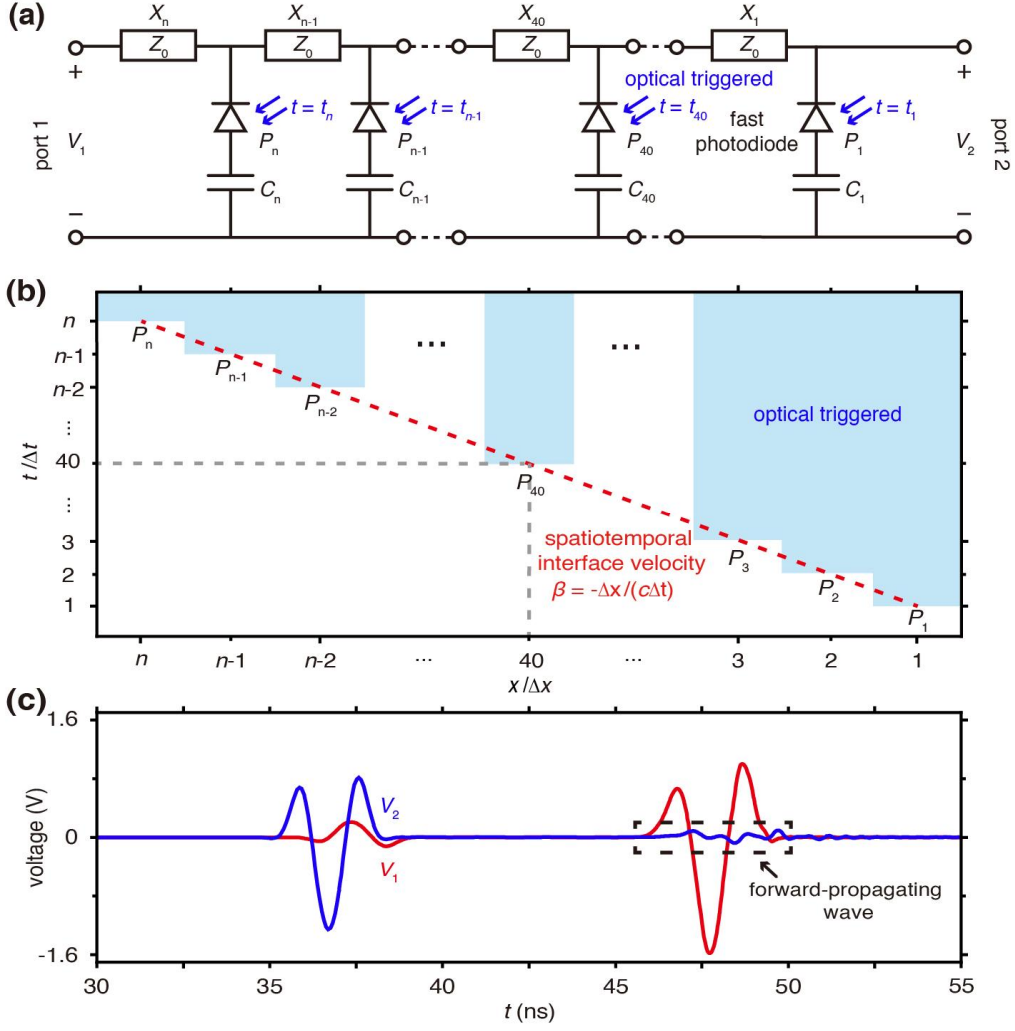

**Figure 5.** Demonstration of generalized coherent wave control in the platform of microstrip transmission lines (MTLs). (a) Equivalent circuit model of spatiotemporal MTLs. Refractive indices of MTLs are controlled by paralleling capacitors, and the on/off state of capacitors is controlled by fast photodiode switches. The parallel capacitance of each MTL is $C = 2\,\text{pF}$. When the photodiode is not triggered (triggered), the refractive index in the vicinity of MTLs is $n_1 = 2.60$ ($n_2 = 3.80$). (b) Triggering scheme of fast photodiodes. Photodiode P$_1$ is triggered at $t = 1\,\text{ns}$, while other photodiodes P$_2$, ... P$_n$ are triggered in sequence with the equivalent time interval of $\Delta t = 1\,\text{ns}$. The effective interface velocity $\beta = -0.0334$. (c) Voltage waves as a function of time. The forward-propagating wave after the interactions between two incidences and the moving interface is highlighted in the black rectangle.

## 5. Conclusion

In this work, we have successfully demonstrated that coherent wave control could be extended to the spatiotemporally engineered media. This extension has greatly relaxed the constraint of identical incident frequencies for the conventional coherent wave control, thus giving rise to various novel phenomena such as flexibly eliminating the forward- or backward-propagating waves, reshaping waveforms with waves in different frequencies, and generating ultrafast pulses using low-frequency pulses. Significantly, we have implemented a detailed simulation of the elimination of spatiotemporal forward-propagating waves on microstrip transmission lines. We highlight that our designs are experimentally feasible, as previous work has already realized a variety of time-varying media in the platforms of, e.g., time-variant transmission lines, time-varying metasurfaces and water surface waves applied with time-dependent voltage.[23, 35-37] In addition, our findings may also inspire future forthcoming research in the field of time-varying medium. For example, how to generalize the coherent perfect absorption from the stationary media into the spacetime-engineered media remains elusive. The spacetime-engineered media with specific levels of loss have the potential to achieve novel coherent perfect absorption, where all the multiple incidences with different frequencies are completely absorbed by the media. Meanwhile, it remains a topic of ongoing investigation to extend to the concept of coherent wave control in systems featuring accelerated interfaces.[38]

**Supporting Information**

**Acknowledgements**
This work was financially supported by the National Natural Science Foundation of China (Grant Nos. 12404363, 12174281, 92050104, 12274314, 61701246), Distinguished Professor Fund of Jiangsu Province (Grant No. 1004-YQR23064), Selected Chinese Government Talent-recruitment Programs of Nanjing (Grant No. 1004-YQR23122), Startup Grant of Nanjing University of Aeronautics and Astronautics (Grant No. 1004-YQR23031), Fundamental Research Funds for the Central Universities (Grant No. NS2023022), and Natural Science Foundation of Jiangsu Province (Grant No. BK20221240).

**Conflict of Interest**
The authors declare no conflict of interest.